\documentclass[12pt]{article}
\usepackage{latexsym,amsfonts,amssymb}
\makeatletter
\@addtoreset{equation}{subsection}
\makeatother

\begin{document}
\begin{center}
{\Large \bf On Cosmological Spacetime Structure and Symmetry:\\
Manifold as a Lie Group, Spinor Structure and Symmetry Group,
Minkowski Metric, and Unnecessariness of Double-Valued
Representations}
\\[1.5cm]
 {\bf Vladimir S.~MASHKEVICH}\footnote {E-mail:
  Vladimir.Mashkevich100@qc.cuny.edu}
\\[1.4cm] {\it Physics Department
 \\ Queens College\\ The City University of New York\\
 65-30 Kissena Boulevard\\ Flushing, New York
 11367-1519} \\[1.4cm] \vskip 1cm

{\large \bf Abstract}
\end{center}

It is shown that cosmological spacetime manifold has the structure
of a Lie group and a spinor space. This leads naturally to the
Minkowski metric on tangent spaces and the Lorentzian metric on
the manifold and makes it possible to dispense with double-valued
representations.

\newpage

\section*{Introduction and Synopsis}

The concept of spacetime symmetry plays a fundamental role in
physics. In particular, the classification of elementary particles
is based on irreducible representations of symmetry groups. When
treating the concept, it is necessary to distinguish between the
symmetry of spacetime manifold, $M$, per se and that of the
manifold equipped with a metric, $(M,g)$.

In General Relativity (GR), which is a local theory, $M=M^{4}$ is
a smooth four-dimensional manifold, and no specific global
structure of $M$ is considered. Metric $g$, being a dynamical
object, is not fixed. Therefore the symmetry group both of $M$ and
of $(M,g)$ is that of diffeomorphisms of $M$.

A conventional approach to the particle classification is based on
the Poincar$\acute{\mathrm{e}}$ group and its subgroups
$SO(3,1)^{\uparrow}=L^{\uparrow}_{+}$ (the restricted Lorentz
group) and $SO(3)$ (the group of rotations in three-dimensional
Euclidean space). In GR, the groups $SO(3)$ and
$SO(3,1)^{\uparrow}$ figure as symmetry groups of the tangent
space $M_{p}$ at a point $p\in M$ equipped with the Minkowski
metric $\eta =\mathrm{diag}(-1,1,1,1)$, which is metric $g$ in
inertial coordinates. Thus it is assumed that $g$ is a Lorentzian
metric, i.e., a metric of signature $(-1,1,1,1)$.

But the true (or proper) representations of $SO(3)$ and
$SO(3,1)^{\uparrow}$ do not include half-integer representations,
which are realized in nature. The situation is revised by
introducing double-valued representations of $SO(3)$ and
$SO(3,1)^{\uparrow}$ or by making use of the universal covering
groups [1] $SU(2)$ and $SL(2,\mathbb{C})$ of $SO(3)$ and
$SO(3,1)^{\uparrow}$, respectively, and their true
representations. Another treatment in GR is based on a spin
structure for $M$.

Now, $SO(3)$ and $SO(3,1)^{\uparrow}$ are symmetry groups of space
and spacetime of Special Relativity (SR) rather than of GR. As to
$SU(2)$ and $SL(2,\mathbb{C})$, they bear no relation to SR space
and spacetime per se. Therefore it is natural to raise the problem
of the symmetry of GR space and spacetime. Since metric is not
fixed, the problem concerns the symmetry of spacetime manifold
$M$.

We start from the cosmological spacetime manifold [2],
$M^{\mathrm{cosm}}=T^{\mathrm{cosm}}\times S^{\mathrm{cosm}}$
where $T^{\mathrm{cosm}}$ is cosmological time and
$S^{\mathrm{cosm}}=S^{3}$ (three-sphere) is cosmological space. A
crucial point is that $S^{3}$ is diffeomorphic to $SU(2)$ [3], so
we put $S^{\mathrm{cosm}}=SU(2)$. A symmetry group of
$S^{\mathrm{cosm}}$ is $SU(2)$ itself, which warrants the choice
of a closed space.

The tangent space $M^{\mathrm{cosm}}_{p}$, $p\in
M^{\mathrm{cosm}}$, is the direct sum of the tangent spaces
$T^{\mathrm{cosm}}_{t}$ and $S^{\mathrm{cosm}}_{s}$,
$M^{\mathrm{cosm}}_{p}=T^{\mathrm{cosm}}_{t}\oplus
S^{\mathrm{cosm}}_{s},\;p=(t,s),\;t\in T^{\mathrm{cosm}},\;s\in
S^{\mathrm{cosm}}$. We have $S^{\mathrm{cosm}}_{s}=su(2)$ [the Lie
algebra of the Lie group $SU(2)$]. A basis for $su(2)$ is
$\{\tau_{j}=\mathrm{i}\sigma_{j}:j=1,2,3\}$ where $\sigma_{j}$ are
the Pauli matrices. The $\tau_{j}$ are anti-Hermitian,
$\tau^{\dag}_{j}=-\tau_{j}$. A basis for $T^{\mathrm{cosm}}_{t}$
is $cI_{1},\;c\in \mathbb{C}$, so that a basis for
$M^{\mathrm{cosm}}_{p}$ is
$\{\tau_{\mu}:\mu=0,1,2,3\},\;\tau_{0}=cI_{2}$. We invoke the
invariance of the basis choice to put $c=\mathrm{i}$, so that
$\tau^{\dag}_{0}=-\tau_{0},\;\tau_{0}=\mathrm{i}\sigma_{0},
\;\sigma_{0}=I_{2}$, and $\tau_{\mu}=\mathrm{i}\sigma_{\mu}$. Thus
$M^{\mathrm{cosm}}_{p}=u(1)\oplus su(2)$ where $u(1)$ is the Lie
algebra of the Lie group $U(1)$.

We have $T^{\mathrm{cosm}}=\mathbb{R}$ with the group
$\mathbb{R}=\{\mathrm{real},+\}$ and a universal covering
homomorphism $\mathbb{R}\rightarrow U(1)$ [4]. Put
$t=(T_{0}/2\pi)\theta$, so that $\mathbb{R}=
\{\theta,+\},\;\;\theta\mapsto e^{\mathrm{i}\theta}$\,;\quad
$M^{\mathrm{cosm}}=\sum_{m=-\infty}^{m=\infty}M^{\mathrm{cosm}(n)}\,,\;\\
M^{\mathrm{cosm}(n)}=U(1)\otimes
SU(2),\;M^{\mathrm{cosm}(n)}\leftrightarrow
\theta_{n}=\bar{\theta}+2\pi n,\;0\leq \bar{\theta}<2\pi$. Thus we
have arrived at the oscillating universe.

The basis $\{\tau_{\mu}=\mathrm{i}\sigma_{\mu}:\mu=0,1,2,3\}$,
leads in a standard way [1], [5] to the Minkowski metric on
$M^{\mathrm{cosm}}_{p}$ and, hence, to the Lorentzian metric on
$M^{\mathrm{cosm}}$.

The $SU(2)$ group structure of $S^{\mathrm{cosm}}$ gives rise to a
spinor structure for the latter. (That structure should not be
confused with the spin structure [3], [5].) We have $SU(2)\ni
a_{SU}\leftrightarrow z:=(z^{1},z^{2})^{\mathrm{tr}}\in Z_{1}$
where $\mathrm{tr}$ stands for transpose,
$|z|:=(|z^{1}|^{2}+|z^{2}|^{2})^{1/2}=1$, and $Z_{1}$ is the space
of the spinors of three-dimensional space [6]. Next we introduce
the space of the spinors of four-dimensional space [6]
$Z_{+}=\{\zeta=(\zeta^{1},\zeta^{2})^{\mathrm{tr}}:|\zeta|>0\}$
and put $\zeta\leftrightarrow (|\zeta|,z),\;z=\zeta/|\zeta|\in
Z_{1},\;|\zeta|=e^{\theta}$. Thus
$Z_{+}\ni\zeta\leftrightarrow(\theta,z)\leftrightarrow(t,s)= p\in
M^{\mathrm{cosm}}$, so that $M^{\mathrm{cosm}}$ has the structure
of the spinor space $Z_{+}$.

Consider operations of the group $GL(2,\mathbb{C})$ on $Z_{+}$. We
have $GL(2,\mathbb{C})\ni
a_{GL}=re^{\mathrm{i}\varphi}a_{SL},\;a_{SL}\in SL(2,\mathbb{C})$.
Operations of
$\{e^{\mathrm{i}\varphi}I_{2}:\varphi\in\mathbb{R}\}$ are reduced
to those of $SU(2)$. Taking this into account, we introduce the
group $PL(2,\mathbb{C})=\{a_{GL}\in
GL(2,\mathbb{C}):\mathrm{det}\,a_{GL}>0\},\quad
PL(2,\mathbb{C})\ni
a_{PL}=ra_{SL},\;r=(\mathrm{det}\,a_{PL})^{1/2}$. We have [7]
$a_{SL}=a_{SU}e^{h},\;h^{\dag}=h,\;\mathrm{Tr}\,h=0$, so that
$a_{PL}=ra_{SL}=ra_{SU}e^{h};\quad SU(2)\subset
SL(2,\mathbb{C})\subset PL(2,\mathbb{C})\supset \{rI_{2}\}$.

Let us summarize: A symmetry group of $M^{\mathrm{cosm}}$ is
$PL(2,\mathbb{C})$: $\zeta\mapsto\zeta'=a_{PL}\zeta;\quad rI_{2}$:
time transformation (time shift),
$\theta\mapsto\theta,\;z'=z;\quad a_{SU}$: space transformation
(space rotation), $\theta'=\theta,\;z\mapsto z';\quad e^{h}$:
spacetime transformation (boost), $\theta\mapsto\theta',\;z\mapsto
z';\quad a_{SL}$: boost and space rotation.

Now that it is $SU(2)$ and $SL(2,\mathbb{C})$, not $SO(3)$ and
$SO(3,1)^{\uparrow}$, that are symmetry groups of spacetime
manifold, there is no need for double-valued representations.

\section{Spacetime, symmetry groups, and double-valued
representations: A conventional treatment}

\subsection{Euclidean space: $SO(3)$ and $SU(2)$}

The Euclidean space has $SO(3)$ as its symmetry group. But the
true representations of $SO(3)$ do not include half-integer
representations. In order to remedy the situation it is necessary
to introduce the so-called double-valued representations of
$SO(3)$ or to make use of the representations of the universal
covering group $SU(2)$ [1], [3].

A homomorphism from $SU(2)$ to $SO(3)$ is constructed as follows
[3]. Introduce a standard basis for the Lie algebra $su(2)$ of the
Lie group $SU(2)$:
\begin{equation}\label{1.1.1}
\{\tau_{j}=\mathrm{i}\sigma_{j}:j=1,2,3\},\quad
\tau^{\dag}_{j}=-\tau_{j},\;\mathrm{Tr}\,\tau_{j}=0
\end{equation}
where the $\sigma_{j}$ are the Pauli matrices:
\begin{equation}\label{1.1.2}
\sigma_{1}=\left(
\begin{array}{cc}
0&1\\ 1&0
\end{array}
\right)\qquad \sigma_{2}= \left(
\begin{array}{cc}
0&-\mathrm{i}\\ \mathrm{i}&0
\end{array}
\right)\qquad \sigma_{3}=\left(
\begin{array}{cc}
1&0\\0&-1
\end{array}
\right)
\end{equation}
Let $\overrightarrow{x}=\{x^{j}:j=1,2,3\},\;x_{j}=x^{j}$, be a
vector of the Euclidean space. Introduce the matrix
\begin{equation}\label{1.1.3}
x_{su}=x^{j}\tau_{j}=\mathrm{i}\left(\begin{array}{cc}
x^{3}&x^{1}-\mathrm{i}x^{2}\\ x^{1}+\mathrm{i}x^{2}&-x^{3}
\end{array}
\right),\quad x_{su}\in su(2),\quad x_{su}\leftrightarrow
\{x^{j}\}
\end{equation}
We have
\begin{equation}\label{1.1.4}
x^{j}y_{j}=\frac{1}{2}[\mathrm{det}(x_{su}+y_{su})-
\mathrm{det}\,x_{su}-\mathrm{det}\,y_{su}],\quad
x^{j}x_{j}=\mathrm{det}\,x_{su}
\end{equation}
Consider a transformation
\begin{equation}\label{1.1.5}
\overrightarrow{x}\mapsto \overrightarrow{x}'
\end{equation}
generated by a transformation
\begin{equation}\label{1.1.6}
x_{su}\mapsto x_{su}'=a_{SU}x_{su}a_{SU}^{\dag}, \quad a_{SU}\in
SU(2),\quad x_{su}'\in su(2)
\end{equation}
Since
\begin{equation}\label{1.1.7}
\mathrm{det}\,x_{su}'=\mathrm{det}\,x_{su},\quad
x'^{j}y'_{j}=x^{j}y_{j}
\end{equation}
we have
\begin{equation}\label{1.1.8}
\overrightarrow{x}'=a_{SO}\overrightarrow{x},\quad a_{SO}\in
SO(3),\qquad \pm a_{SU}\leftrightarrow a_{SO}
\end{equation}
which gives rise to the double-valued representations of $SO(3)$.
In particular, a spinor of three-dimensional space is transformed
into its negative when that space undergoes one complete rotation.

Consider [6]
\begin{equation}\label{1.1.9}
a_{SU}(\varphi)=\left(\begin{array}{cc}e^{-\mathrm{i}\varphi/2}&0\\
0&e^{\mathrm{i}\varphi/2}\end{array}\right)\qquad a_{SO}(\varphi)=
\left(\begin{array}{ccc}\cos\varphi&-\sin\varphi&0\\
\sin\varphi&\cos\varphi&0\\ 0&0&1\end{array}\right)
\end{equation}
with
\begin{equation}\label{1.1.10}
\pm a_{SU}(\varphi)\leftrightarrow a_{SO}(\varphi)
\end{equation}
We have
\begin{equation}\label{1.1.11}
a_{SU}(\varphi+2\pi)=-a_{SU}(\varphi),\quad a_{SO}(\varphi+2\pi)=
a_{SO}(\varphi)
\end{equation}
so that
\begin{equation}\label{1.1.12}
(a_{SU}(\varphi),a_{SU}(\varphi+2\pi))\leftrightarrow
a_{SO}(\varphi)
\end{equation}
In particular,
\begin{equation}\label{1.1.13}
a_{SO}(2\pi)=a_{SO}(0)=I_{3},\quad
a_{SU}(2\pi)=-a_{SU}(0)=-I_{2},\; a_{SU}(4\pi)=a_{SU}(0)=I_{2}
\end{equation}

\subsection{Minkowski spacetime: $SO(3,1)^{\uparrow}$ and $SL(2,\mathbb{C})$}

In the case of the Minkowski spacetime, the situation is similar
to that of the Euclidean space. A symmetry group is
$SO(3,1)^{\uparrow}=L^{\uparrow}_{+}$ (the restricted Lorentz
group), the universal covering group is $SL(2,\mathbb{C})$. A
homomorphism from $SL(2,\mathbb{C})$ to $SO(3,1)^{\uparrow}$ is
constructed as follows [1], [5]. Introduce a standard basis for
the Lie algebra $u(2)$:
\begin{equation}\label{1.2.1}
\{\tau_{\mu}=\mathrm{i}\sigma_{\mu}:\mu=0,1,2,3\}, \quad
\tau_{\mu}^{\dag}=-\tau_{\mu}
\end{equation}
\begin{equation}\label{1.2.2}
\sigma_{0}=I_{2}=\left(\begin{array}{cc} 1&0\\0&1
\end{array}\right)
\end{equation}
Let $x=\{x^{\mu}\},\;x_{0}=-x^{0},\;x_{j}=x^{j}$, be a vector of
the Minkowski spacetime. Introduce the matrix
\begin{equation}\label{1.2.3}
x_{u}=x^{\mu}\tau_{\mu}=\mathrm{i}\left(\begin{array}{cc}
x^{0}+x^{3}&x^{1}-\mathrm{i}x^{2}\\x^{1}+\mathrm{i}x^{2}&x^{0}-x^{3}
\end{array}\right),\quad x_{u}\in u(2),\quad x_{u}\leftrightarrow x
\end{equation}
We have
\begin{equation}\label{1.2.4}
x^{\mu}y_{\mu}=\frac{1}{2}[\mathrm{det}(x_{u}+y_{u})-
\mathrm{det}\,x_{u}-\mathrm{det}\,y_{u}],\quad
x^{\mu}x_{\mu}=\mathrm{det}\,x_{u}
\end{equation}
Consider a transformation
\begin{equation}\label{1.2.5}
x\mapsto x'
\end{equation}
generated by a transformation
\begin{equation}\label{1.2.6}
x_{u}\mapsto x_{u}'=a_{SL}x_{u}a_{SL}^{\dag},\quad a_{SL}\in
SL(2,\mathbb{C}),\quad x_{u}'\in u(2)
\end{equation}
Since
\begin{equation}\label{1.2.7}
\mathrm{det}\,x_{u}'=\mathrm{det}\,x_{u},\quad
x'^{\mu}y'_{\mu}=x^{\mu}_{\mu}y_{\mu}
\end{equation}
we have
\begin{equation}\label{1.2.8}
x'=a_{L_{+}^{\dag}}x,\quad a_{L_{+}^{\dag}}\in L_{+}^{\dag},
\qquad \pm a_{SL}\leftrightarrow a_{L_{+}^{\dag}}
\end{equation}
which gives rise to the double-valued representations of
$SO(3,1)^{\dag}=L_{+}^{\dag}$. In particular, the Dirac field is
transformed into its negative when three-dimensional space
undergoes one complete rotation.

\subsection{Drawback}

A conventional invocation of the groups $SU(2)$ and
$SL(2,\mathrm{C})$ for the purpose of obtaining half-integer
representations suffers from an obvious drawback: It has an
unnatural character since both $SU(2)$ and $SL(2,\mathbb{C})$ per
se bear no relation to the Euclidean space and the Minkowski
spacetime. In this connection we quote some authors:

[3]: ``Note that in quantum mechanics, starting from the
commutation relations of the angular momentum operators, we only
know the Lie algebra $so(3)\cong su(2)$. Experiments tell us that
not only the rotation group $SO(3)$ but also its universal
covering group $SU(2)$ are represented in nature: Some
interference experiments with neutrons\ldots do distinguish
between $g$ and $-g$ in $SU(2)$.''

[1]: ``In application to physical systems possessing rotational
symmetry, there is no {\it a priori} reason to decide whether the
double-valued representations occur in nature or not. In reality,
we know that they do exist---all fermion (i.e. half-odd-integer
spin) systems, such as electrons, protons,\ldots etc., are
described by quantum mechanical wave functions that correspond to
double-valued representations of $SO(3)$. Of course, the
single-valued representations are also realized in nature---they
correspond to boson systems (with integer spin).''

[5]: ``Here we examine the physicist's somewhat paradoxical
statement that the Dirac electron field is transformed into its
negative when space undergoes one complete rotation.''

[8]: ``It is sometimes said that the $j$ half-integral
representations of $SU(2)$ are `double-valued' representations of
$SO(3)$, allowed because of the nature of the measurement process.
We prefer to think of it differently. The $j$ half-integral
representations are faithful representations of $SU(2)$. Moreover,
$SO(3)$ is a homomorphic image of $SU(2)$, and thus some of the
representations of $SU(2)$ are not representations of $SO(3)$.

The heart of the matter is that we are accustomed to thinking in
terms of either a geometric three-dimensional space $R_{3}$ or a
four-dimensional spacetime $R_{4}$. All our concepts of geometry
derive ultimately from the strong coupling of our senses (eyes)
with photons (spin 1). We therefore interact strongly with the
$\mathcal{D}^{1}$ representations---the half-integral
representations are unfamiliar. If our eyes were constructed to
interact with $j=1/2$ particles, we would see a peculiar twofold
degeneracy in those properties of the universe depending on
integral spin particles.''

\section{Cosmological spacetime manifold}

\subsection{Direct product manifold}

Cosmological spacetime manifold, $M^{\mathrm{cosm}}$, is the
direct product manifold [2]:
\begin{equation}\label{2.1.1}
M^{\mathrm{cosm}}=T^{\mathrm{cosm}}\times S^{\mathrm{cosm}},\quad
M^{\mathrm{cosm}}\ni p=(t,s),\;t\in T^{\mathrm{cosm}},\;s\in
S^{\mathrm{cosm}}
\end{equation}
where $T^{\mathrm{cosm}}$ is cosmological time and
$S^{\mathrm{cosm}}$ is cosmological space. Cosmological time
\begin{equation}\label{2.1.2}
T^{\mathrm{cosm}}=\mathbb{R}
\end{equation}
Cosmological space is the three-sphere (the closed universe):
\begin{equation}\label{2.1.3}
S^{\mathrm{cosm}}=S^{3}=\{x_{k}:k=1,2,3,4,\;\sum_{k}x_{k}^{2}=1\}
\end{equation}

\subsection{Tangent space}

The tangent space of $M^{\mathrm{cosm}}$ at a point $p\in
M^{\mathrm{cosm}},\;M_{p}^{\mathrm{cosm}}$, is the direct sum of
the tangent spaces of $T^{\mathrm{cosm}}$ and $S^{\mathrm{cosm}}$:
\begin{equation}\label{2.2.1}
M^{\mathrm{cosm}}_{p}=T^{\mathrm{cosm}}_{t}\oplus
S^{\mathrm{cosm}}_{s},\quad p=(t,s)
\end{equation}

\section{Group structure: Manifold as a Lie group}

\subsection{Space as $SU(2)$}

A crucial point is that $S^{3}$ is diffeomorphic to the Lie group
$SU(2)$ [3]. We have
\begin{equation}\label{3.1.1}
SU(2)\ni a_{SU}=\left(\begin{array}{cc}z^{1}&- z^{2\ast}\\z^{2}&
z^{1\ast}
\end{array}\right),\quad |z^{1}|^{2}+|z^{2}|^{2}=1
\end{equation}
Put
\begin{equation}\label{3.1.2}
z^{1}=x_{1}+\mathrm{i}x_{2},\;\;z^{2}=x_{3}+\mathrm{i}x_{4},\quad
|z^{1}|^{2}+|z^{2}|^{2}=\sum_{k}x_{k}^{2}=1
\end{equation}
so that
\begin{equation}\label{3.1.3}
SU(2)\ni a_{SU}\leftrightarrow (z^{1},z^{2})\leftrightarrow
\{x_{k}\}\in S^{3}
\end{equation}
So we put
\begin{equation}\label{3.1.4}
S^{\mathrm{cosm}}=SU(2)
\end{equation}
A symmetry group of $SU(2)$ is $SU(2)$ itself---by a left action
[3]:
\begin{equation}\label{3.1.5}
a_{SU}\mapsto a'_{SU}=\overline{a}_{SU}a_{SU}
\end{equation}

Contemporary  cosmology is faced with the problem of choice
between a flat and a closed space. A conventional approach to the
problem is based on continuous properties of the space [9], [10].
But a certain choice is possible only on the basis of a discrete
characteristic of the latter. A symmetry group is just the one.
Representations realized in nature are those of $SU(2)$, not only
of $SO(3)$. Therefore it is the closed space that should be
preferred.

\subsection{Tangent space of space: $su(2)$,
Euclidean metric, $SO(3)$ and $SU(2)$ }

The tangent space of $SU(2)$ at the point $I_{2}\in SU(2)$ is
$su(2)$ [the Lie algebra of the Lie group $SU(2)$]:
\begin{equation}\label{3.2.1}
S^{\mathrm{cosm}}_{I_{2}}=su(2)
\end{equation}
Now the results of Subsection 1.1 follow for the tangent
space---as a natural consequence of (3.2.1).

\subsection{Tangent space of spacetime: $u(1)\oplus su(2)$}

A basis for the tangent space $T^{\mathrm{cosm}}_{t}$ is
\begin{equation}\label{3.3.1}
cI_{1}=c,\quad c\in \mathbb{C}
\end{equation}
so that a basis for $M^{\mathrm{cosm}}_{p}$ (2.2.1),
$p=(t,s=I_{2})$, is
\begin{equation}\label{3.3.2}
\{cI_{2},\tau_{1},\tau_{2},\tau_{3}\},\quad
\tau_{j}=\mathrm{i}\sigma_{j}
\end{equation}
We have $\tau^{\dag}_{j}=-\tau_{j}$, and to retain this property
under a change of the basis we put
\begin{equation}\label{3.3.3}
c=\mathrm{i}
\end{equation}
so that
\begin{equation}\label{3.3.4}
T^{\mathrm{cosm}}_{t}=u(1)\;\;(\mathrm{the\; Lie\; algebra\; of\;
the\; Lie\; group}\;U(1))
\end{equation}
and
\begin{equation}\label{3.3.5}
M^{\mathrm{cosm}}_{p=(t,I_{2})}=u(1)\oplus su(2)
\end{equation}

Now the results of Subsection 1.2 follow for the tangent
space---as a natural consequence of (3.3.5).

\subsection{Spacetime manifold as $U(1)\otimes SU(2)$
and the oscillating universe}

We have $T^{\mathrm{cosm}}=\mathbb{R}$ (2.1.2) with the additive
group structure of $\mathbb{R}$ and a universal covering
homomorphism [4]
\begin{equation}\label{3.4.1}
\mathbb{R}\rightarrow U(1)
\end{equation}
Put
\begin{equation}\label{3.4.2}
T^{\mathrm{cosm}}\ni t=\frac{T_{0}}{2\pi}\theta
\end{equation}
so that
\begin{equation}\label{3.4.3}
\mathbb{R}=\{\theta,+\},\quad \mathbb{R}\ni \theta\mapsto
e^{\mathrm{i}\theta}\in U(1)
\end{equation}

Now, taking into account (2.1.1), we put
\begin{equation}\label{3.4.4}
M^{\mathrm{cosm}}=\sum_{n=-\infty}^{\infty}M^{\mathrm{cosm}(n)},
\quad M^{\mathrm{cosm}(n)}\leftrightarrow
\theta^{(n)}=\overline{\theta}+2\pi n,\quad
0\leq\overline{\theta}<2\pi
\end{equation}
\begin{equation}\label{3.4.5}
M^{\mathrm{cosm}(n)}=U(1)\otimes SU(2)
\end{equation}
Thus we have arrived at the oscillating universe with the Lie
group structure (3.4.5) of spacetime manifold.

\subsection{Tangent bundle structure}

Let us turn our attention to a global structure of the set of
tangent spaces of $M^{\mathrm{cosm}(n)}$, i.e., of the tangent
bundle. The Lie algebra structure is given by the space of
left-invariant vector fields [4], which forms a Lie algebra under
the commutator of vector fields. The Minkowski spacetime structure
is given by the Minkowski frame bundle [3], [5],
$F^{\mathrm{Minkowski}}(M^{\mathrm{cosm}(n)})$. Since
$M^{\mathrm{cosm}(n)}$ is parallelizable [3], the Minkowski frame
bundle is trivial:
\begin{equation}\label{3.5.1}
F^{\mathrm{Minkowski}}(M^{\mathrm{cosm}(n)})=M^{\mathrm{cosm}(n)}\times
L^{\uparrow}_{+}
\end{equation}
A section is
\begin{equation}\label{3.5.2}
\sigma(p)=(p,\{\tau_{\mu}(p):\mu=0,1,2,3\})
\end{equation}
where
\begin{equation}\label{3.5.3}
\begin{array}{l}
\tau_{\mu}(p)=\Lambda^{\nu}{}_{\mu}(p)\tau_{\nu}=
a_{SL}(p)\tau_{\mu}a_{SL}(p)^{\dag},\quad \Lambda(p)\in
L_{+}^{\dag},\;\;a_{SL}\in SL(2,\mathbb{C})\\
\Lambda(p)\leftrightarrow \pm a_{SL}(p)
\end{array}
\end{equation}
and
\begin{equation}\label{3.5.4}
\begin{array}{l}
\tau_{0}(p)=\tau_{0},\;\;\tau_{j}(p)=R^{l}{}_{j}(p)\tau_{l}=
a_{SU}(p)\tau_{j}a_{SU}(p)^{\dag},\quad R(p)\in
SO(3),\;\;a_{SU}\in SU(2)\\ R(p)\leftrightarrow \pm a_{SU}(p)
\end{array}
\end{equation}
In $F^{\mathrm{Minkowski}}(M^{\mathrm{cosm}(n)})$, the fiber over
$p\in M^{\mathrm{cosm}(n)}$ is the set of all Minkowski bases of
$M^{\mathrm{cosm}(n)}_{p}$, i.e., bases with metric
$\mathrm{diag}(-1,1,1,1)$.

The above treatment gives rise to the spin structure for
$M^{\mathrm{cosm}(n)}$ [3], [5].

\subsection{Lorentzian metric}

In GR, the Minkowski metric $\mathrm{diag}(-1,1,1,1)$ is a metric
$g$ in inertial coordinates. Although $g$ is a dynamical object,
which is not fixed, its signature is fixed. Thus we arrive at the
conclusion that $g$ should be a Lorentzian metric.

\section{Spinor structure}

\subsection{Spinor structure of space:
Spinors of three-dimensional space}

The $SU(2)$ group structure of cosmological space
$S^{\mathrm{cosm}}$ (3.1.4) gives rise to a spinor structure for
$S^{\mathrm{cosm}}$. (A spinor structure should not be confused
with the spin structure.) Introduce the space of the spinors of
three-dimensional space [6]:
\begin{equation}\label{4.1.1}
Z_{1}=\{z:=(z^{1},z^{2})^{\mathrm{tr}}:z^{A}\in\mathbb{C},
\;|z|:=(|z^{1}|^{2}+|z^{2}|^{2})^{1/2}=1\}
\end{equation}
where $\mathrm{tr}$ stands for transpose.

The space $Z_{1}$ is equipped with the Hermitian product [11]
\begin{equation}\label{4.1.2}
(z|z')=\sum_{A}^{1,2}z^{A\ast}z'^{A}
\end{equation}
and with the invariant scalar product [11]
\begin{equation}\label{4.1.3}
zCz'=z^{A}C_{AB}z'^{B},\quad A,B=1,2
\end{equation}
where
\begin{equation}\label{4.1.4}
C=\left(\begin{array}{cc}0&1\\-1&0
\end{array}\right)
\end{equation}
Both the Hermitian and invariant scalar products are invariant
under transformations of the group $SU(2)$:
\begin{equation}\label{4.1.5}
z\mapsto z'=a_{SU}z,\quad a_{SU}\in SU(2)
\end{equation}
\begin{equation}\label{4.1.6}
(a_{SU}z|a_{SU}z')=(z|z')
\end{equation}
\begin{equation}\label{4.1.7}
a_{SU}zCa_{SU}z'=zCz'
\end{equation}

In view of (3.1.1), we have
\begin{equation}\label{4.1.8}
SU(2)\ni a_{SU}\leftrightarrow z\in Z_{1}
\end{equation}
so that $S^{\mathrm{cosm}}$ has the structure of $Z_{1}$.

\subsection{Spinor structure of spacetime:
Spinors of four-dimensional space}

Now we introduce the space of the spinors of four-dimensional
space [6]
\begin{equation}\label{4.2.1}
Z_{+}=\{\zeta:=(\zeta^{1},\zeta^{2})^{\mathrm{tr}}:
\zeta^{A}\in\mathbb{C},\;|\zeta|:=(|\zeta^{1}|^{2}+
|\zeta^{2}|^{2})^{1/2}>0\}
\end{equation}
and the space of dotted spinors [6] or cospinors [12]
\begin{equation}\label{4.2.2}
\dot{Z}_{+}=\{\dot{\zeta}:=
(\zeta^{\dot{1}},\zeta^{\dot{2}})^{\mathrm{tr}}:
\zeta^{\dot{A}}\in\mathbb{C},\;|\dot{\zeta}|>0\}
\end{equation}
(In the physical literature, $\zeta$ and $\dot{\zeta}$ are called
a right-handed and a left-handed Weyl spinors, respectively [13].)

In addition to the Hermitian product
\begin{equation}\label{4.2.3}
(\eta|\xi)=\sum_{A}^{1,2}\eta^{A\ast}\xi^{A}
\end{equation}
and the invariant scalar product
\begin{equation}\label{4.2.4}
\eta C\xi=\eta^{A}C_{AB}\xi^{B}
\end{equation}
there is the scalar product [12]
\begin{equation}\label{4.2.5}
(\dot{\eta}|\xi)\in\mathbb{C}
\end{equation}
which is nondegenerate in both $\xi$ and $\dot{\eta}$ and linear
in $\xi$ and antilinear in $\dot{\eta}$:
\begin{equation}\label{4.2.6}
(\dot{\eta}|c_{1}\xi+c_{2}\zeta)=
c_{1}(\dot{\eta}|\xi)+c_{2}(\dot{\eta}|\xi),\;\;
(c_{1}\dot{\eta}+c_{2}\dot{\zeta}|\xi)=c_{1}^{\ast}(\dot{\eta}|\xi)
+c_{2}^{\ast}(\dot{\zeta}|\xi),\quad c_{k}\in\mathbb{C}
\end{equation}
The scalar product is invariant under these transformations:
\begin{equation}\label{4.2.7}
\xi\mapsto \xi'=a_{GL}\xi\leftrightarrow \dot{\xi}\mapsto
\dot{\xi}'=(a_{GL}^{\dag})^{-1}\dot{\xi},\quad a_{GL}\in
GL(2,\mathbb{C})
\end{equation}
The invariant scalar product is invariant under transformations of
the group $SL(2,\mathbb{C})$:
\begin{equation}\label{4.2.8}
\xi\mapsto\xi'=a_{SL}\xi,\quad a_{SL}\in SL(2,\mathbb{C})
\end{equation}
The Hermitian product is invariant under transformations of the
group $SU(2)$:
\begin{equation}\label{4.2.9}
\xi\mapsto\xi'=a_{SU}\xi,\quad a_{SU}\in SU(2)
\end{equation}

Now introduce the correspondence
\begin{equation}\label{4.2.10}
\zeta\leftrightarrow(|\zeta|,z),\quad z=\zeta/|\zeta|\in Z_{1}
\end{equation}
and put
\begin{equation}\label{4.2.11}
  |\zeta|=e^{\theta},\quad \zeta\leftrightarrow(\theta,z)
\end{equation}
Thus we have
\begin{equation}\label{4.2.12}
Z_{+}\ni \zeta\leftrightarrow(\theta,z)\leftrightarrow(t,s)=p\in
M^{\mathrm{cosm}}
\end{equation}
so that $M^{\mathrm{cosm}}$ has the structure of the spinor space
$Z_{+}$. Specifically,
\begin{equation}\label{4.2.13}
|\zeta|\leftrightarrow\theta\leftrightarrow t\in
T^{\mathrm{cosm}}:\;\mathrm{time}
\end{equation}
\begin{equation}\label{4.2.14}
z\leftrightarrow a_{SU}\leftrightarrow s\in
S^{\mathrm{cosm}}:\;\mathrm{space}
\end{equation}
\begin{equation}\label{4.2.15}
\zeta\leftrightarrow(\theta,z)\leftrightarrow(t,s)\in
T^{\mathrm{cosm}}\times S^{\mathrm{cosm}}:\;\mathrm{spacetime}
\end{equation}

\section{Symmetry group}

\subsection{Operations of $GL(2,\mathbb{C})$}

Consider operations of the group $GL(2,\mathbb{C})$ on the spinor
space $Z_{+}$. We have
\begin{equation}\label{5.1.1}
GL(2,\mathbb{C})\ni a_{GL}=re^{\mathrm{i}\varphi}a_{SL},\quad
a_{SL}\in SL(2,\mathbb{C})
\end{equation}
so that
\begin{equation}\label{5.1.2}
a_{GL}\zeta=re^{\mathrm{i}\varphi}a_{SL}\zeta,\;\;a_{SL}\zeta=:
\overline{\zeta}=|\overline{\zeta}|\overline{z},\;\;
a_{GL}\zeta=r|\overline{\zeta}|e^{\mathrm{i}\varphi}\overline{z}
\end{equation}
Now
\begin{equation}\label{5.1.3}
z\leftrightarrow a_{SU}\in SU(2)
\end{equation}
Consider
\begin{equation}\label{5.1.4}
a_{SU}'\left(\begin{array}{cc}z^{1}&-z^{2\ast}\\z^{2}&z^{1\ast}
\end{array}\right)=\left(\begin{array}{cc}
e^{\mathrm{i}\varphi}z^{1}&-(e^{\mathrm{i}\varphi}z^{2})^{\ast}\\
e^{\mathrm{i}\varphi}z^{2}&(e^{\mathrm{i}\varphi}z^{1})^{\ast}
\end{array}\right)=:a_{SU}^{\varphi}
\end{equation}
i.e.,
\begin{equation}\label{5.1.5}
a_{SU}'a_{SU}=a_{SU}^{\varphi},\quad
a_{SU}'=a_{SU}^{\varphi}(a_{SU})^{-1}
\end{equation}
We have
\begin{equation}\label{5.1.6}
a_{SU}'\left(\begin{array}{l}z^{1}\\z^{2}\end{array}\right)=
\left(\begin{array}{l}e^{\mathrm{i}\varphi}z^{1}\\
e^{\mathrm{i}\varphi}z^{2}\end{array}\right),\quad
a_{SU}'z=e^{\mathrm{i}\varphi}z
\end{equation}
Thus operations of $\{e^{\mathrm{i}\varphi}I_{2}\}$ are reduced to
those of $SU(2)$. (See also [6].)

\subsection{$PL(2,\mathbb{C})$: Symmetry group of manifold}

Taking into account the results proved above, introduce the group
\begin{equation}\label{5.2.1}
PL(2,\mathbb{C})=\{a_{GL}\in GL(2,\mathbb{C}):
\mathrm{det}\,a_{GL}>0\}
\end{equation}
We have
\begin{equation}\label{5.2.2}
PL(2,\mathbb{C})\ni
a_{PL}=ra_{SL},\;\;r=(\mathrm{det}\,a_{PL})^{1/2},\;\;a_{SL}\in
SL(2,\mathbb{C})
\end{equation}
A symmetry group of $M^{\mathrm{cosm}}$ is $PL(2,\mathbb{C})$:
\begin{equation}\label{5.2.3}
\zeta\mapsto\zeta'=a_{PL}\zeta
\end{equation}

\subsection{Operations of subgroups of $PL(2,\mathbb{C})$}

We have [7]
\begin{equation}\label{5.3.1}
a_{SL}=a_{SU}e^{h},\;\;h^{\dag}=h,\;\;\mathrm{Tr}\,h=0
\end{equation}
so that
\begin{equation}\label{5.3.2}
a_{PL}=ra_{SL}=ra_{SU}e^{h}
\end{equation}
Consider operations of subgroups of $PL(2,\mathbb{C})$
\begin{equation}\label{5.3.3}
SU(2)\subset SL(2,\mathbb{C})\subset
PL(2,\mathbb{C})\supset\{rI_{2}\}
\end{equation}
on $M^{\mathrm{cosm}}$:
\begin{equation}\label{5.3.4}
rI_{2}:\; \mathrm{time\;transformation\;(time\;shift)},\;\;
\theta\mapsto\theta',\;z'=z
\end{equation}
\begin{equation}\label{5.3.5}
a_{SU}:\;\mathrm{spase\;transformation\;
(space\;rotation)},\;\;\theta'=\theta,\;z\mapsto z'
\end{equation}
\begin{equation}\label{5.3.6}
\begin{array}{l}
e^{h}:\;\mathrm{spacetime\;transformation\;(boost)},\;\;
\theta\mapsto\theta',\;z\mapsto z'\\ T\times S\rightarrow T'\times
S'\\ T^{\mathrm{cosm}}\times
S^{\mathrm{cosm}}\;\mathrm{is\;distinguished\;in\;connection\;with\;
dynamics}
\end{array}
\end{equation}
\begin{equation}\label{5.3.7}
a_{SL}:\;\mathrm{boost\;and\;space\;rotation}
\end{equation}
\begin{equation}\label{5.3.8}
a_{PL}:\;\mathrm{boost,\;space\;rotation,\;and\;time\;shift}
\end{equation}

\subsection{Farewell to double-valued representations}

Since it is $SU(2)$ and $SL(2,\mathrm{C})$, not $SO(3)$ and
$SO(3,1)^{\uparrow}=L^{\uparrow}_{+}$, that are symmetry groups of
the spacetime manifold, there is no need for double-valued
representations. It is single-valued, or true representations that
are realized in nature.

\section*{Acknowledgments}

I would like to thank Alex A. Lisyansky for support and Stefan V.
Mashkevich for helpful discussions.

\end{document}